\documentclass[10pt,a4paper]{article}

\usepackage[affil-it]{authblk} % for affiliation

\usepackage{graphicx} % for figures

\usepackage{caption} % for captions

\usepackage{natbib} % for biliography

\providecommand{\keywords}[1]
{
	\small	
	\textbf{\textit{Keywords---}} #1
} % for keywords

\usepackage{amsmath} % for symbols

\newcommand{\adj}[1]{{#1}_{adj}}
\newcommand{\avg}{\mathrm{E}}
\newcommand{\var}{\mathrm{Var}}
\newcommand{\prob}{\mathrm{Pr}}
\newcommand{\quant}{\mathrm{q}}
\newcommand{\lik}{\mathcal{L}}
\newcommand{\llik}{\ell}
\newcommand{\clik}{\lik_C}
\newcommand{\score}{u}
\newcommand{\vest}{\mathrm{V}}
\newcommand{\nest}{\mathrm{N}}
\newcommand{\dOR}{\rho}
\newcommand{\avgdiff}{\delta}
\newcommand{\trace}{\mathrm{trace}}

\title{Adjusted composite likelihood \\ for robust Bayesian meta-analysis}
\author{Michele Lambardi di San Miniato}
\author{Nicola Sartori}
\affil{Department of Statistical Sciences, University of Padova, \\ Via Cesare Battisti 241, 35121 Padova, Italy}
\date{2021-03-20}

\begin{document}

\maketitle

\begin{abstract}
	A composite likelihood is a non-genuine likelihood function that allows to make inference on limited aspects of a model, such as marginal or conditional distributions. Composite likelihoods are not proper likelihoods and need therefore calibration for their use in inference, from both a frequentist and a Bayesian perspective. The maximizer to the composite likelihood can serve as an estimator and its variance is assessed by means of a suitably defined sandwich matrix. In the Bayesian setting, the composite likelihood can be adjusted by means of magnitude and curvature methods. Magnitude methods imply raising the likelihood to a constant, while curvature methods imply evaluating the likelihood at a different point by translating, rescaling and rotating the parameter vector. Some authors argue that curvature methods are more reliable in general, but others proved that magnitude methods are sufficient to recover, for instance, the null distribution of a test statistic. We propose a simple calibration for the marginal posterior distribution of a scalar parameter of interest which is invariant to monotonic and smooth transformations. This can be enough for instance in medical statistics, where a single scalar effect measure is often the target.
\end{abstract}


\section{Introduction}

Likelihood functions are popular tools both in frequentist and Bayesian inference. A genuine, or proper, likelihood is defined in terms of a probability distribution, evaluated at the observed full dataset and treated as a function of the parameter only. In the Bayesian framework, under a suitable prior, a genuine likelihood implies a posterior distribution for the parameter that can be used in inference. A composite likelihood is defined by multiplying two or more genuine likelihood components together, but generally this product is not a genuine likelihood in turn. This happens because the product implies a working independence assumption that may not be valid in the light of background knowledge. If each of the genuine likelihood components is correctly specified, the maximum composite likelihood estimator is at least consistent under a suitable asymptotic theory. Moreover, the estimation variance can be assessed via a suitably defined sandwich covariance matrix. For more on the theory of composite likelihoods in the classical framework, one can refer to the review by \cite{Varin}.

As to the Bayesian framework, a composite likelihood is not based on the density of a correctly specified model, so it represents a wrong probability, and the inverse probabilities implied by Bayes' theorem are incorrect too: this may imply either liberal or conservative inference, as first warned by \cite{Monahan}, depending for instance on the sign of neglected correlations among other things. For more on diagnostic methods to assess this type of behavior, one can refer to \cite{Lazar}.

Consider the case of clustered data. In a rather simple case, the observations will just make up data pairs. These pairs will be independent of each other and internally correlated in general. This way of modeling can be useful in clinical meta-analyses, where the treatment and control group are paired within the studies and the studies are independent. A composite likelihood can help to infer a marginal or conditional distribution, without the need to assume an explicit form for the internal correlations. One can refer for instance to \cite{Chandler} on the topic of composite likelihoods for clustered data.

\begin{figure}[h]
	\centering
	\caption{\label{fig:chendata}}
	\includegraphics[width=\linewidth]{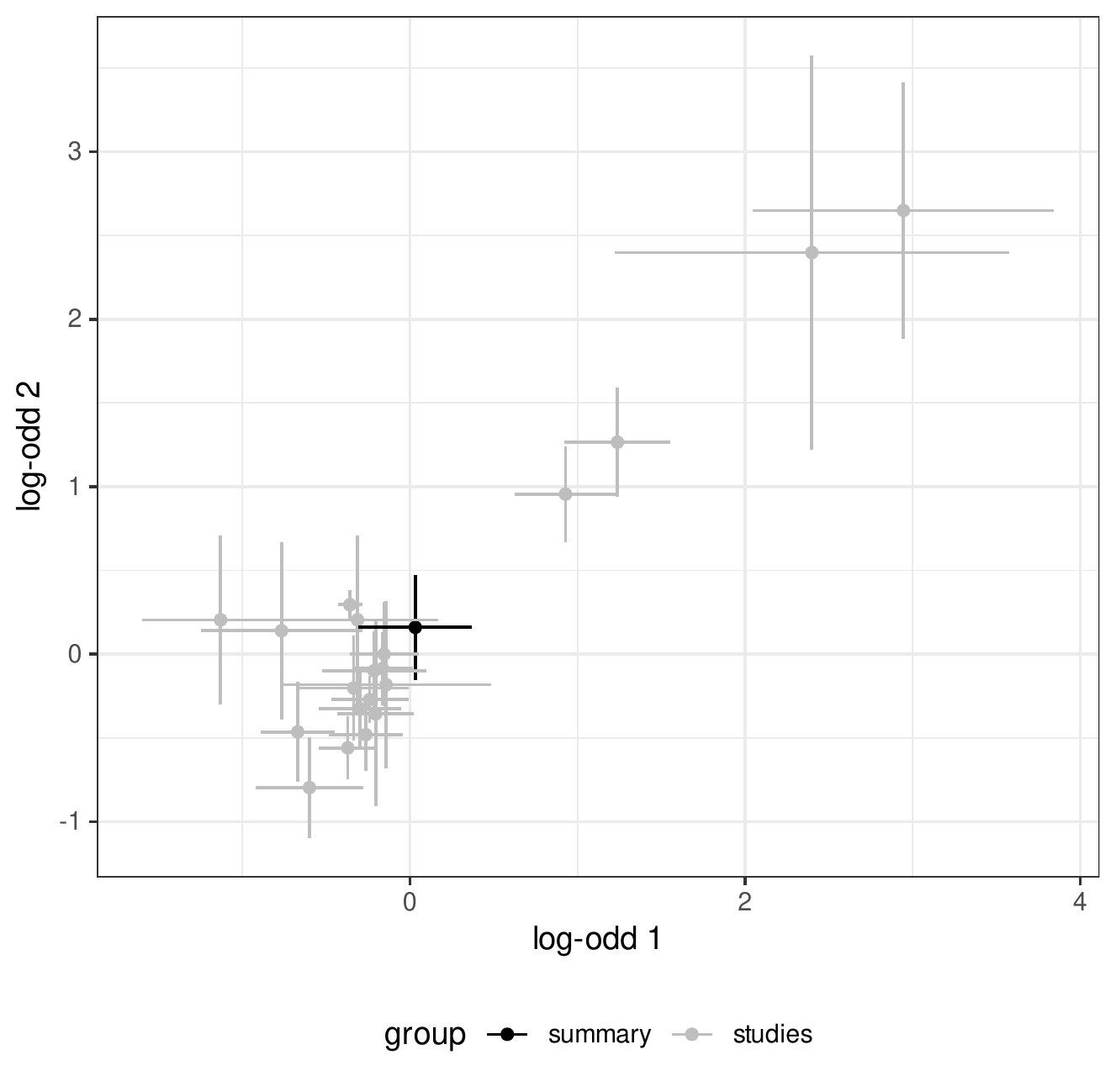}
\end{figure}

\cite{Chen} consider the case of a meta-analysis on clustered count data, where the counts of patients experiencing a given event are reported among two treatment groups, say 1 and 2, for each study. Even if the groups are independent by randomization, a spurious correlation between groups and within studies can be observed as a result of heterogeneity between the studies. See for instance Figure \ref{fig:chendata}, where we report the logarithm of event odds along with standard errors for each treatment group and study. We also computed the summary log-odds, along with standard errors estimated in a robust fashion accounting for correlations. The studies may be heterogeneous just by involving patients with different baseline frailties: this is reflected in the Figure by the presence of a covariant gradient among log-odds.

Modeling the spurious correlation is not of substantial interest and it essentially a side-effect of meta-analysis. \cite{Chen} resort to a marginal composite likelihood, for which they specify simply the marginal distribution for the outcome in each group. In order to assess estimation variance they only have to assume the existence of the spurious correlation within studies. They can only make inference on functions of marginal parameters, which are actually their sole interest. This approach clearly saves from making potentially wrong assumptions that are also difficult to formulate, compare, and check in practice.

One can compare this approach with an alternative, also considered by \cite{Chen}, which involves explicit modeling of correlations within studies. The most flexible approach in this sense is probably the copula modeling approach, for which one may refer for instance to \cite{Nelsen}. It is difficult though to find an appropriate copula model in every situation, as there might also be too little studies for a meta-analysis, and this kind of approach will likely require a model selection step. This is actually avoided by means of a composite likelihood approach.

As to the Bayesian framework, a composite likelihood will usually require an adjustment, or modification, that essentially preserves the composite's definition, while correctly accounting for uncertainty. Posterior distributions could be mis-calibrated otherwise, as first warned by \cite{Monahan} for the more general case of pseudo likelihoods. Adjustments are generally motivated on the basis of asymptotic theory, as in \cite{Miller}. The aim is to approximately recover some nice feature of genuine posterior distributions, in the logic illustrated by \cite{Stoehr}. \cite{Ribatet} review two main classes of adjustments broadly referred to as curvature and magnitude methods. As they make further distinctions between overall and adaptive adjustments, we will only deal with overall adjustments, which are simpler because they require less tuning. Set aside their specific motivations, these methods can still be tested in simulations, where the ground truth is known by design and several datasets are simulated and analyzed automatically. Calibration plots can be used to check the validity of probabilistic statements based on the adjusted posterior distributions, in the fashion illustrated by \cite{Lazar}.

The most flexible adjustment is the curvature method, which involves rescaling and rotating the parameter vector, in order to make posterior variance close to the sandwich variance from the frequentist framework. The log-posterior is asymptotically quadratic, so this is sufficient to recover approximate calibration for all parameters. This method is advocated for by \cite{Ribatet} when the analysis has no specific target but potentially aims at making inference on the whole parameter vector.

\cite{Pauli} aim to recover the null chi-squared distribution of a composite likelihood ratio test. In doing this, they first recall Bartlett corrections, which essentially multiply the log-likelihood by a constant. This constant is also called temperature and the adjustment itself is also known as tempered composite likelihood. This operation results into a magnitude adjustment, which can be enough to target specific aspects of the posterior distribution, but it is still not flexible and it will not work under reparameterization. The authors proposing it deem the method as targeted to the whole parameter vector, so we will refer to it as an \textit{omnibus} adjustment.

In this paper we argue that the \textit{omnibus} magnitude adjustment used by \cite{Pauli} is less safe than claimed, and we show this fact in the case of an independence composite likelihood. Curvature methods advocated for by \cite{Ribatet}, for instance, constitute a safer alternative in general, but in many applications the aim can be to just make inference on a scalar parameter of interest: in such a case, we propose a simple \textit{targeted} magnitude adjustment that calibrates the marginal posterior distribution of specifically that parameter of interest. Our adjustment is also invariant to transformations of the target parameter.

We compare the omnibus magnitude adjustment with our targeted proposal and with the curvature adjustment. As matrix square roots are involved in the latter, we compare a definition of such square root that is common in applications with an alternative definition proposed by \cite{Kessy}: this proposal reduces the implied rotation of the parameter vector implied by curvature adjustments and preserves better some features of the composite likelihood, including its asymmetries, skewness, kurtosis and so forth. A simulation study on a clustered data model with an independence likelihood will show that a magnitude adjustment can suffice in making inference about a scalar parameter of interest: in this sense our adjustment seems as safe as a curvature, though limited \textit{a priori} to the target parameter, while \cite{Pauli}'s version of magnitude adjustment implies no substantial correction because of some mathematical subtleties that will be illustrated later.

In medical statistics and other disciplines, where some effect measures approximate each other, targeting a given parameter may also target approximately a related one, making the adjustment reliable to a wider extent. If there is a focus, targeted magnitude adjustments can be as good performers as the curvature method, while remaining unreliable in general for all parameters but the target.

\section{Composite likelihood}

Let $\theta$ be a $p$-dimensional parameter. Let $y=(y_1,\dots,y_K)$ be an observed outcome of the data-generating process with true data distribution $f_0(y)$. The outcome is modeled by a random vector $Y=(Y_1,\dots,Y_K)$. We assume that one is willing to formulate $P_1(\cdot;\theta),\dots,P_K(\cdot;\theta)$, the marginal distributions of $Y_1,\dots,Y_K$, respectively, whereas no interest lies on the joint distributions. We assume that at least these marginals are correct, so there exists a true parameter $\theta_0$ such that
$$P_k(y_k;\theta_0)=\int f_0(y) \, d y_{-k}$$
Then, $K$ genuine likelihoods $\lik_1,\dots,\lik_K$ can be defined, as follows:
$$\lik_1(\theta)=P_1(y_1;\theta) \,,\quad \dots \,,\quad \lik_K(\theta)=P_K(y_K;\theta) \,.$$
These definition can hold up to a proportionality constant. A composite likelihood $\clik(\theta)$ can be defined as their product
$$\clik(\theta) = \prod_{k=1}^K \lik_k(\theta) \,.$$
One may read $\clik(\theta)$ in the expression above as a measure of the probability to observe $y$ under $\theta$: then, the product in the definition of $\clik(\theta)$ implies the assumption that the random vectors $Y_1,\dots,Y_K$ are all independent. This is referred to as a working independence assumption, which can be wrong without affecting the consistency of estimators.

Let $\partial$ denote the gradient or Jacobian operator. We also need to define the following likelihood-related quantities. Notation is mostly in line with the review on composite likelihoods offered by \cite{Varin}.

$\llik_k(\theta)=\log\lik_k(\theta)$ is the generic $k$-th log-likelihood.

$\score_k(\theta)=\partial\llik_k(\theta)/\partial\theta$ is the generic $k$-th score function.

$\score(\theta)=\sum_{k=1}^K \score_k(\theta)$ is the score function.

$\hat{H}_k(\theta)=-\partial\score_k(\theta)$ is the generic $k$-th sample sensitivity matrix.

$H_k(\theta)=\avg\left\{\hat{H}_k(\theta);\theta\right\}$ is the generic $k$-th sensitivity matrix.

$H(\theta)=\sum_{k=1}^K H_k(\theta)$ is the sensitivity matrix.

$J(\theta) = \var\left\{\score(\theta);\theta\right\}$ is the variability matrix.

$\hat{J}(\theta)$ is a sample variability matrix that serves as a consistent estimator of $J(\theta)$.

After a suitable definition of sample size $n$, under regularity conditions, the maximum composite likelihood estimator $\hat{\theta}$ maximizing the composite likelihood is a consistent estimator of $\theta$. The variance of $\hat{\theta}$ can be approximated by the robust asymptotic variance $\vest_\theta$, defined as
$$\vest_\theta = H(\theta)^{-1} J(\theta) H(\theta)^{-1} \,.$$
This is called robust, as contrasted to the naive variance $\nest_\theta$, which is defined as
$$\nest_\theta = H(\theta)^{-1} \,.$$
The latter provides the variance assessment when trusting the composite likelihood as if it were genuine. The robust and naive variances are defined as the inverse of Godambe's and Fisher's information matrices, respectively.

In practice, $\vest_\theta$ and $\nest_\theta$ can be estimated feasibly with
$$\hat{\vest}_\theta=\hat{H}(\hat{\theta})^{-1} \hat{J}(\hat{\theta}) \hat{H}(\hat{\theta})^{-1} \quad\text{and}\quad \hat{\nest}_\theta=\hat{H}(\hat{\theta})^{-1} \,.$$

Assume a scalar parameter $\eta=\eta(\theta)$ is of interest, with $\eta(\cdot)$ suitably smooth. Let $\partial\eta$ be the gradient of $\eta$ with respect to $\theta$. The maximum composite likelihood estimator is equivariant, so $\hat{\eta}=\eta(\hat{\theta})$. Moreover, $\hat{\theta}$ is consistent and has bias of order $O(n^{-1})$, and the same is true for $\hat{\eta}$. The robust variance of $\hat{\eta}$, denoted by $\vest_\eta$, can be approximated using the delta method, with
$$\vest_\eta=\partial\eta^\top \cdot \vest_\theta \cdot \partial\eta \,.$$
We call this the robust delta variance of $\hat{\eta}$. A naive version can be defined as
$$\nest_\eta = \partial\eta^\top \cdot \nest_\theta \cdot \partial\eta \,.$$
Estimating the naive and the robust variance for $\eta$ also involves evaluating $\partial\eta$ at $\hat{\eta}$.

\subsection{Clustered data case}

The definition of $\hat{J}(\theta)$ can be based on background knowledge about the true correlation structure, which is neglected under the working independence assumption when defining the composite likelihood. Consider a meta-analysis summarizing $N$ studies: each study implies a cluster of observations; the clusters are independent of each other, but they are to be treated as internally correlated so to account for heterogeneity-related issues.

One can define $Y_k=(Y_{1k},\dots,Y_{ik},\dots,Y_{Nk})$, for $k=1,\dots,K$, where the studies are indexed by $i=1,\dots,N$ and the random variable $Y_{ik}$ models the outcome $y_{ik}$ from the $k$-th treatment group of the $i$-th study.

The following likelihood-related quantities can be defined.

$P_{ik}(\cdot;\theta)$ is the distribution of $Y_{ik}$,

$\lik_{ik}(\theta)=P_{ik}(y_{ik};\theta)$ is the $i$-th study's contribution to $\lik_k(\theta)$, so $\lik_k(\theta) = \prod_{i=1}^N \lik_{ik}(\theta)$,

$\llik_{ik}(\theta)=\log\lik_{ik}(\theta)$ is the $i$-th study's contribution to $\llik_k(\theta)$, so $\lik_k(\theta) = \sum_{i=1}^N \llik_{ik}(\theta)$,

$\score_{ik}(\theta)=\partial \llik_{ik}(\theta)$ is the $i$-th study's contribution to $\score_k(\theta)$, so $\score_k(\theta) = \sum_{i=1}^N \score_{ik}(\theta)$,

$\score_{i*}(\theta)=\sum_{k=1}^K \score_{ik}(\theta)$ is the $i$-th study's contribution to $\score(\theta)$, so $\score(\theta) = \sum_{i=1}^N \score_{i*}(\theta)$.

Just by assuming the correlation structure, the following holds:
$$J(\theta) = \var\left\{\score(\theta);\theta\right\} = \var\left\{\sum_{i=1}^N \score_{i*}(\theta);\theta\right\} = \sum_{i=1}^N\var\left\{\score_{i*}(\theta);\theta\right\} \,.$$
So, by treating $\score_{i*}(\theta)$ as a column vector, an estimator for $J(\theta)$ can be defined as
$$\hat{J}(\theta) = \sum_{i=1}^N \score_{i*}(\theta) \cdot \score_{i*}(\theta)^\top \,.$$

\section{Bayesian adjustment}

The composite likelihood approach has clear advantages over full likelihoods in the case we consider because there could hardly be any knowledge about the appropriate copula model for the problem at hand. The composite likelihood does not need to explicitly formulate such uninteresting aspects of the model.

Composite likelihoods are more naturally dealt with in the classical framework than the Bayesian one. \cite{Monahan} discussed the risks of mis-calibrated posterior distributions when dealing with pseudo, not genuine, likelihoods, such as the pseudo likelihood from Cox' proportional hazards model. A composite likelihood may be replaced with an artificial likelihood in the Bayesian framework, as proposed by \cite{Muller}: the artificial likelihood is Gaussian, with mean vector equal to the maximum composite likelihood estimator $\hat{\theta}$ and covariance matrix equal to the robust variance $\vest_\theta$. This artificial likelihood is also called the sandwich likelihood and it can also be motivated under the asymptotic theory developed by \cite{Miller}.

A composite likelihood is not expected to be well calibrated and therefore needs to be adjusted, or slightly modified, before using it in the Bayesian framework. The adjustment's tuning can be based on the output of a classical analysis and the resulting adjusted composite likelihood should imply an asymptotically calibrated posterior distribution. A reasonable adjustment method should be defined so that one approximately falls back to the composite likelihood when the working independence assumption is correct. Another requirement is that the adjustment should preserve, as far as possible, the shape and the asymmetries of the original composite likelihood: the sandwich likelihood by \cite{Muller} is not particularly suited on this aspect.

The main types of adjustments discussed by \cite{Ribatet} are curvature and magnitude methods. The curvature method transforms the parameter vector before feeding it to the composite likelihood, while the magnitude transforms the composite likelihood itself. The magnitude transformation raises the likelihood to a constant or, equivalently, multiplies the log-likelihood by the same constant. The curvature transformation is affine or linear. Actually, the magnitude method involves just one tuning constant, as contrasted to the curvature method, which involves a tuning matrix made up of $O(p^2)$ constants. This is reflected in the higher flexibility of curvature method with respect to magnitude methods.

\subsection{Curvature adjustment}

The curvature method replaces the composite likelihood $\clik$ with the curvature-adjusted composite likelihood $\adj\clik$, defined as
$$\adj\clik(\theta;A) = \clik(A(\theta-\hat{\theta})+\hat{\theta}) \,.$$
In the expression above, $A$ is a $p \times p$ tuning matrix. In the terms proposed by \cite{Stoehr}, it can be seen as a device that allows the adjusted composite likelihood to satisfy a matching rule stated as follows:
$$\nest_\theta(\adj{\clik}(\cdot;A)) = \vest_\theta \,.$$
Here, $\nest_\theta(\adj{\clik})$ is the naive variance matrix for $\theta$, as implied by the adjusted composite likelihood $\adj{\clik}$. This aspect of $\adj{\clik}$ and the extreme point $\hat{\theta}$ are the only features of $\adj{\clik}$ relevant to the asymptotic theory as developed, for instance, by \cite{Miller}. The matrix $A$ that feasibly solves the problem above is in the form:
$$A =\left\{(\hat{\nest}_\theta^{-1})^{1/2}\right\}^{-1} (\hat{\vest}_\theta^{-1})^{1/2} \,.$$
Here, $B=A^{1/2}$ denotes the square root of matrix $A$, such that $B^\top B = A$. As a matter of linear algebra, the matrix square root is not unique: for every orthonormal matrix $E$, also $E\cdot B$ can serve as a square root of $A$. This phenomenon is called rotational freedom. Different definitions of the square root imply different optimality or invariance properties, as discussed by \cite{Kessy}.

In simulations we considered a main version of curvature adjustment based on a definition of matrix square root referred to as zero-phase component analysis (ZCA) in \cite{Kessy}. This definition is popular among practitioners, even under a different name, since it has the same eigen-decomposition as the argument matrix excepted the eigenvalues, which are square-rooted. ZCA is also symmetric whenever the argument is, and it is endowed with the nice property
$$(A^{1/2})^{-1} = (A^{-1})^{1/2} \,.$$
We also considered an experimental variant of curvature adjustment, involving the ZCA-cor definition of matrix square root, which is invariant to rescaling operations when the argument is a covariance matrix. This alternative implied no substantial improvement in simulations but it is reported for reference. This should be a concern in general, though, because different matrix square roots imply different rotations that can more or less alter the asymmetries of the composite likelihood. Such a rotation effect does not matter asymptotically, but it can do in finite samples.

\cite{Shaby} proposed an adjustment similar to curvature method in some respect, but it first draws samples from the unadjusted posterior distribution and then scales and rotates these samples around $\hat{\theta}$ so that its covariances will reflect the robust variance $\vest_\theta$. This solution meets the asymptotic requirements for \cite{Miller} but it reduces the contribution of the prior to the analysis. It will not be discussed further in this paper, though it could be a useful alternative to curvature adjustments from an objective Bayesian perspective.

\subsection{Magnitude adjustments}

The curvature-adjusted posterior is approximately calibrated for all parameters and thus also for any target $\eta$, but this solution is excessively flexible if just $\eta$ is of interest. One may consider an adjustment that calibrates for $\eta$ without calibrating for other parameters, especially if this approach requires less constants to be tuned, resulting into improved stable behavior.

One can magnitude-adjust the composite likelihood by raising it to a suitable power $w$, called temperature, in the following fashion.
$$\adj\clik(\theta;w)=\clik(\theta)^w \,.$$
The term magnitude refers to the change in the order of magnitude of the relative log-likelihood at any given $\theta$ implied by $w$.

This formulation arises naturally for instance in the work by \cite{Pauli}, as a Bartlett correction, as their goal is to recover the null distribution of some likelihood ratio test. In particular, they set
$$w=\frac{p}{\trace\left\{(\nest_\theta)^{-1}\vest_\theta\right\}} \,,$$
where the unary $\trace$ operator defined for square-sized matrices returns the sum of the argument's diagonal entries. This is referred to as the \textit{omnibus} magnitude adjustment in the following. In general, even with a different aim, the tuning constant $w$ allows for less corrections than implied by curvature adjustment's tuning matrix $A$, but this is actually not a problem if the goal is suitably limited by some specific inferential needs.

Under the same logic seen before, as outlined by \cite{Stoehr}, consider a magnitude-adjusted composite likelihood $\adj{\clik}$ that satisfies a different matching rule, which is less demanding and stated as follows:
$$\nest_\eta(\adj{\clik}(\cdot;w)) = \vest_\eta \,.$$
Here, $\nest_\eta(\adj{\clik})$ is the naive variance for $\eta$ alone, as implied by the adjusted composite likelihood $\adj{\clik}$. The tuning constant $w$ that feasibly solves the problem above is:
$$w=\frac{\nest_\eta}{\vest_\eta} \,.$$
The composite log-likelihood is thus rescaled in order to imply the robust variance at least for the target parameter $\eta$. The posterior will be approximately calibrated for $\eta$ after \cite{Miller}. The proposed magnitude adjustment is based on delta variances, so it is the same by design for all smooth invertible transformations of $\eta$.

\section{Marginal beta-binomial likelihood}

\cite{Chen} consider the case of a meta-analysis with independence composite likelihood on count data: the number of patients experiencing a given event of interest is reported for each treatment group in each study. The most natural choice is to model the counts as binomial-distributed, but over-dispersion is often observed in practice, as a result of heterogeneity among studies.

It is convenient to resort to hierarchical modeling. Conditional to the probability of event $p_{ik}$ in the $k$-th group of the $i$-th study, $Y_{ik}$ is binomial-distributed with probability of event $p_{ik}$ and known size $n_{ik}$, and it is independent of all other observable outcomes in the light of randomization. If there were no heterogeneity, it would hold $p_{ik}=\pi_k$ for all $i=1,\dots,N$, but the probabilities can otherwise be assumed random, with expected value $\pi_k$ independent of the study.

As detailed by \cite{Chen}, a marginal Beta distribution with shape parameters $\alpha_k,\beta_k$ can be assumed for $p_{ik}$, so the average probability of event in the $k$-th treatment group be defined as
$$\avg(p_{ik};\theta)= \pi_k = \frac{\alpha_k}{\alpha_k+\beta_k} \,,$$
where $\theta=(\alpha_1,\beta_1,\dots,\alpha_K,\beta_K)$. Beta models can flexibly represent random quantities bounded between $0$ and $1$. This choice has some other computational advantages since it implies that, marginally, thus without conditioning on $p_{ik}$, $Y_{ik}$ is beta-binomial-distributed. In formulas, it holds
$$P_{ik}(y;\theta)=p(y;n_{ik},\alpha_k,\beta_k)=\binom{n_{ik}}{y}\frac{B(y+\alpha_k,n_{ik}-y+\beta_k)}{B(\alpha_k,\beta_k)} \,,$$
where $B(\cdot,\cdot)$ is the beta function.

Conditional on probabilities, the counts can be assumed independent due to randomization, but the observed correlations within studies are accounted for in terms of a correlation among latent probabilities $p_{i1},\dots,p_{iK}$. This correlation does not need to be specified for the composite likelihood approach to be used. Actually, the approach is useful in that it does not require to specify that aspect of the model. One can imagine though that there exists a copula model that rules the probabilities identically within each study, resulting in beta marginals for the probabilities. After Sklar's theorem, as per \cite{Nelsen}, such a copula exists and is unique, because the beta distribution is absolutely continuous.

An average odd of event in the $k$-th treatment group can be defined as
$$o_k=\frac{p_k}{1-p_k} \,.$$
\cite{Chen} consider the case where $K=2$ and they aim at making inference on a scalar parameter of interest called diagnostic odds ratio ($\dOR$), which is a function of the marginal parameters defined as:
$$\dOR=\frac{o_1}{o_2} \,.$$
This quantity generalizes the odds ratio when probabilities are random.

In our study, we also consider an alternative effect measure, defined as
$$\avgdiff=p_1-p_2 \,.$$
The two measures $\dOR$ and $\avgdiff$ are related and legitimate alternatives as an effect measure of interest. Our concern is whether an adjusted composite likelihood that is targeted at $\dOR$ is also reliable for $\avgdiff$. This may not be the case in switching to the Bayesian framework, where the composite likelihood can be magnitude-adjusted in order to recover calibration for potentially few aspects of the model.

\section{Calibration plots}

Beyond the rationales, different adjustments can be compared in terms of approximate recovery of posterior calibration: this is the property of a posterior distribution that yields asymptotically valid probabilistic statements. In repeated simulations for which the ground truth is known, the posterior quantile of order $p$ for a given parameter can be compared with the true value of the parameter, which should lie under the posterior quantile with frequency about $p$. The importance of these concerns will naturally depend on the way of interpreting probability. From a classical perspective, \cite{Miller} predicts an asymptotic agreement between classical and Bayesian analyses, so a total disagreement between the two might be concerning.

\cite{Lazar} dealt with empirical likelihoods and assessed posterior calibration for a single scalar parameter of interest. The same approach can be applied to composite likelihood functions, as the aim is the same in checking for posterior calibration. In simulations, the ground truth $\theta_0$ and so $\eta_0=\eta(\theta_0)$ are known, and some properties of the posterior $\prob(\cdot \mid Y=y)$ can be assessed. To this end, \cite{Lazar} defined a statistic, which we will denote $h$. The statistic $h$ for $\eta$ is defined as
$$h=\prob(\eta \leq \eta_0 \mid Y=y) \,.$$
Under repeated sampling, asymptotically it holds
$$h \,\dot{\sim}\, \mathcal{U}\mathrm{niform} [0,1] \,.$$

We define the statistic $g$ as
$$g=|2h-1| \,,$$
which has the same uniform limiting distribution. We use quantile-based credible intervals (QB) instead of highest-posterior-density intervals (HPD) for the sake of simplicity and of invariance to monotonic transformations of $\eta$. If the credibility level is set equal to $p$, the QB interval for $\eta$ is delimited by $\quant(\eta;p/2 \mid Y)$ and $\quant(\eta;1-p/2 \mid Y)$. Here, $\quant(\eta;p\mid Y)$ is $\eta$'s posterior quantile of order $p$, such that
$$\prob\left\{\eta \leq \quant(\eta;p\mid Y)\mid Y\right\} = p \,.$$

In this paper we call calibration plot the empirical cumulative distribution function of statistic $g$. This chart can be read in terms of the effective coverage, along the $y$-axis, as a function of the nominal coverage, along the $x$-axis, of QB credible intervals. The ideal behavior for the curve is a straight line connecting points $(0,0)$ and $(1,1)$, which would imply perfect calibration. Curves above this line denote conservative inference, while any curve below would denote liberal inference. % This plot loses information about the down-ward or up-ward nature of bias in Bayesian estimation of $\eta$, but consistent nature of the classical estimator is not disputed here: mostly, the part of precision assessment is questioned in this kind of low-dimensional problems.

One must consider that the ideal behavior may not be achieved in finite samples, due to non-Gaussian likelihood or informative priors to begin with, so the performance of curvature methods will set the standard in the light of the arguments made by \cite{Ribatet}.

\section{Simulation study}

In simulations, we generate 1000 datasets for each of some selected settings. Every setting is a joint configuration of copula model, its implied correlation, and a phase, all defined as follows.

The copula model, rules the latent probabilities within studies, and ranges in Clayton, Frank, Gumbel copulas. The copula model implies a rank correlation among probabilities within studies, ranging from $0.5$ to about $0.95$ in our simulations; it is set by suitably tuning the copula's only parameter.

Some popular parametric copula models were chosen. These are available in the \texttt{R} package \texttt{copula}\nocite{Rcopula}. Under these models, the family's only parameter determines rank correlations in a identifiable way, independent of marginal distributions. This relation is also exploited by \cite{Tsukahara} for semi-parametric estimation of copulas, because it does not require the estimation of marginal distributions.

%%% added reference for R package copula %%%

The marginal parameters are
$$\theta=(\log\alpha_1,\log\beta_1,\log\alpha_2,\log\beta_2) \,,$$
so the gradients of $\dOR$ and $\avgdiff$ to compute delta variances are
$$\frac{\partial\dOR}{\partial\theta} = \frac{1}{\dOR} \cdot(1,-1,-1,1) \quad\text{and}\quad \frac{\partial\avgdiff}{\partial\theta} = \left[\pi_1\cdot(1,-1),\pi_2\cdot(-1,1)\right] \,.$$
The logarithm is used in the definition of $\theta$ as the posterior is supposed to be better approximated by a Gaussian when the parameters can take any real value. This choice is useful in the Metropolis-Hastings sampling of the posterior distribution, of which we use an adaptive variant implemented in the \texttt{R} package \texttt{MHadaptive}\nocite{RMHadaptive}. The same parameterization is also used in the \texttt{R} package \texttt{xmeta} for the assessment of estimation variance\nocite{Rxmeta}. Burn in and thinning were used to reduce chains' autocorrelation and improve on mixing, which made the simulations computationally burdening.

The sample sizes $n_{ik}$ for each treatment group and study are the same from the meta-analysis dataset used by \cite{Chen}, also available is \texttt{R} package \texttt{xmeta}\nocite{Rxmeta}, while the data $y_{ik}$ were simulated.

As to phases, in phase 1 each dataset is generated conditional to marginal parameters, which are randomly drawn for each dataset from an informative prior for $\theta$. The prior chosen is an uncorrelated Gaussian distribution with zero means and unit variances. In phase 2, the true marginal parameters are fixed and they can take, equally likely, any of the two following configurations:
$$\exp(\theta')=(3.11, 2.91, 3.94, 3.36) \,,\quad \exp(\theta'')=(0.5, 0.5, 0.5, 0.5) \,.$$
These configurations are the same used in experiments by \cite{Chen}. The prior in phase 2 is only used in the Bayesian analysis, and not to draw $\theta$, and it is the same prior as in phase 1 but made uninformative by setting all marginal variances equal to $10^4$ instead of $1$.

The configuration $\theta'$ is chosen on purpose to be close to the actual estimates obtained on the original data by \cite{Chen}. The concern is whether one can consistently infer the truth at least in a neighborhood of some relevant parameter configuration, since the parameter space cannot be probed exhaustively.

\subsection{Results}

\begin{figure}
	\centering
	\caption{\label{fig:calib_a1}}
	\hspace*{-0.3\linewidth}\includegraphics[width=1.6\textwidth,keepaspectratio]{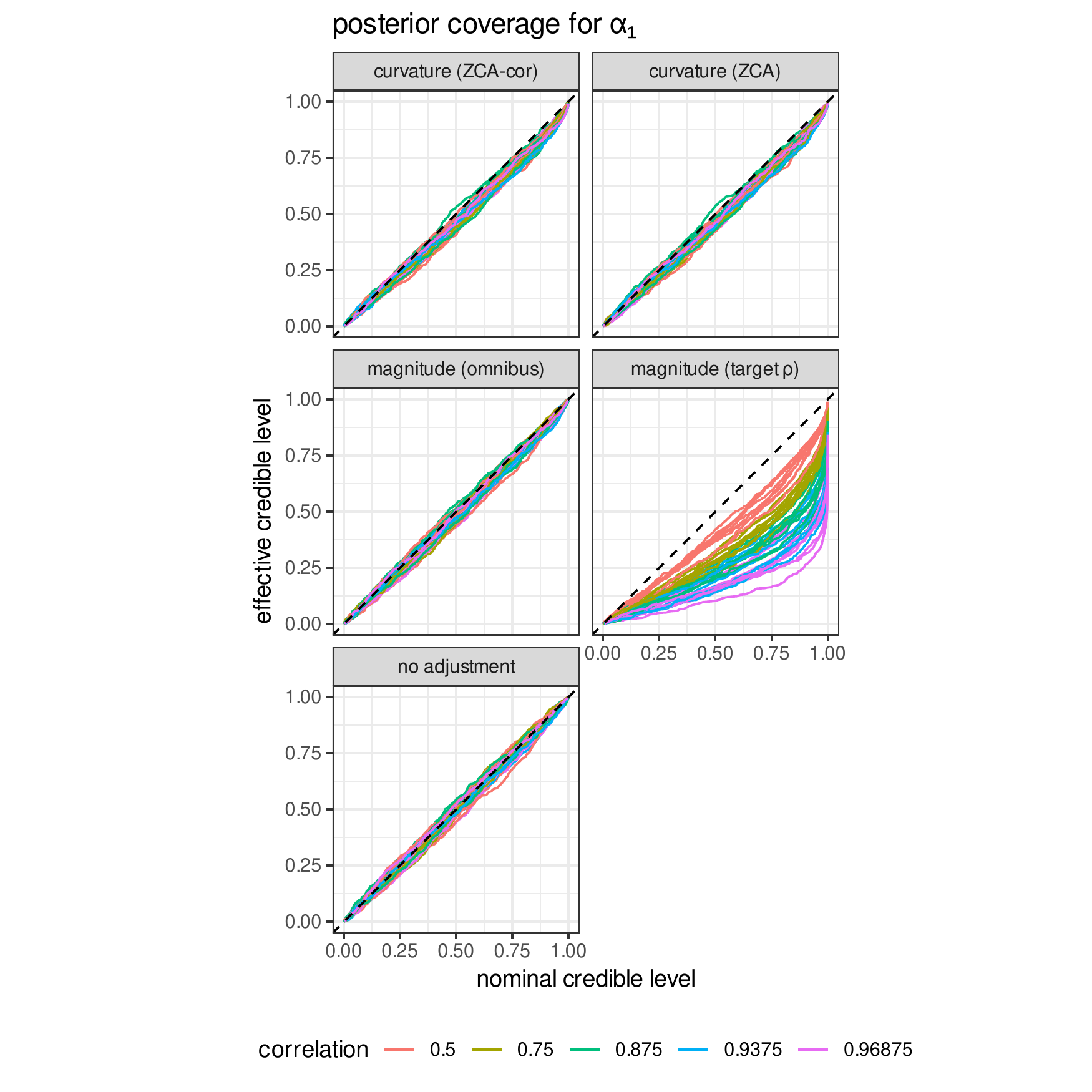}
\end{figure}

\begin{figure}
	\centering
	\caption{\label{fig:calib_dOR}}
	\hspace*{-0.3\linewidth}\includegraphics[width=1.6\textwidth,keepaspectratio]{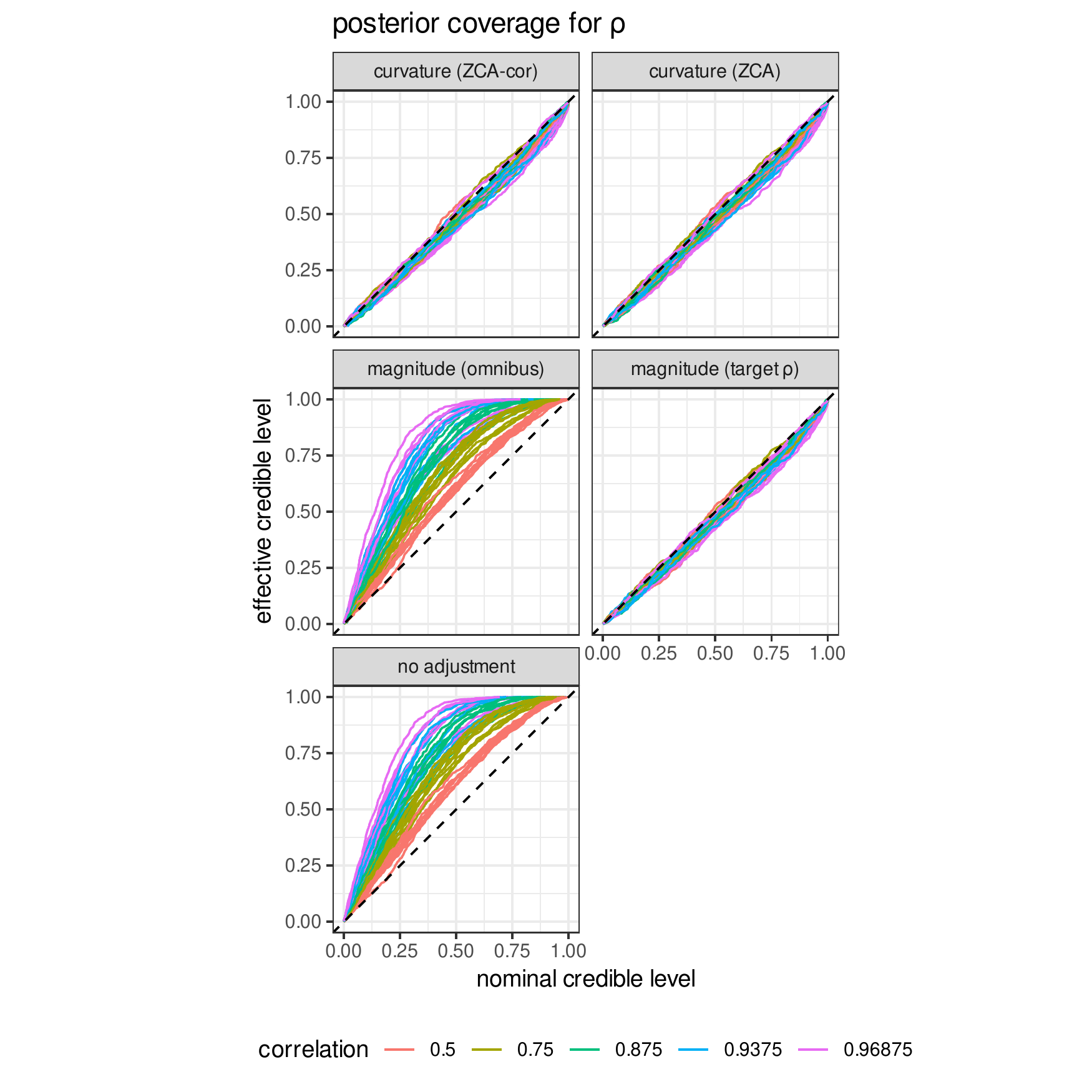}
\end{figure}

\begin{figure}
	\centering
	\caption{\label{fig:calib_avgdiff}}
	\hspace*{-0.3\linewidth}\includegraphics[width=1.6\textwidth,keepaspectratio]{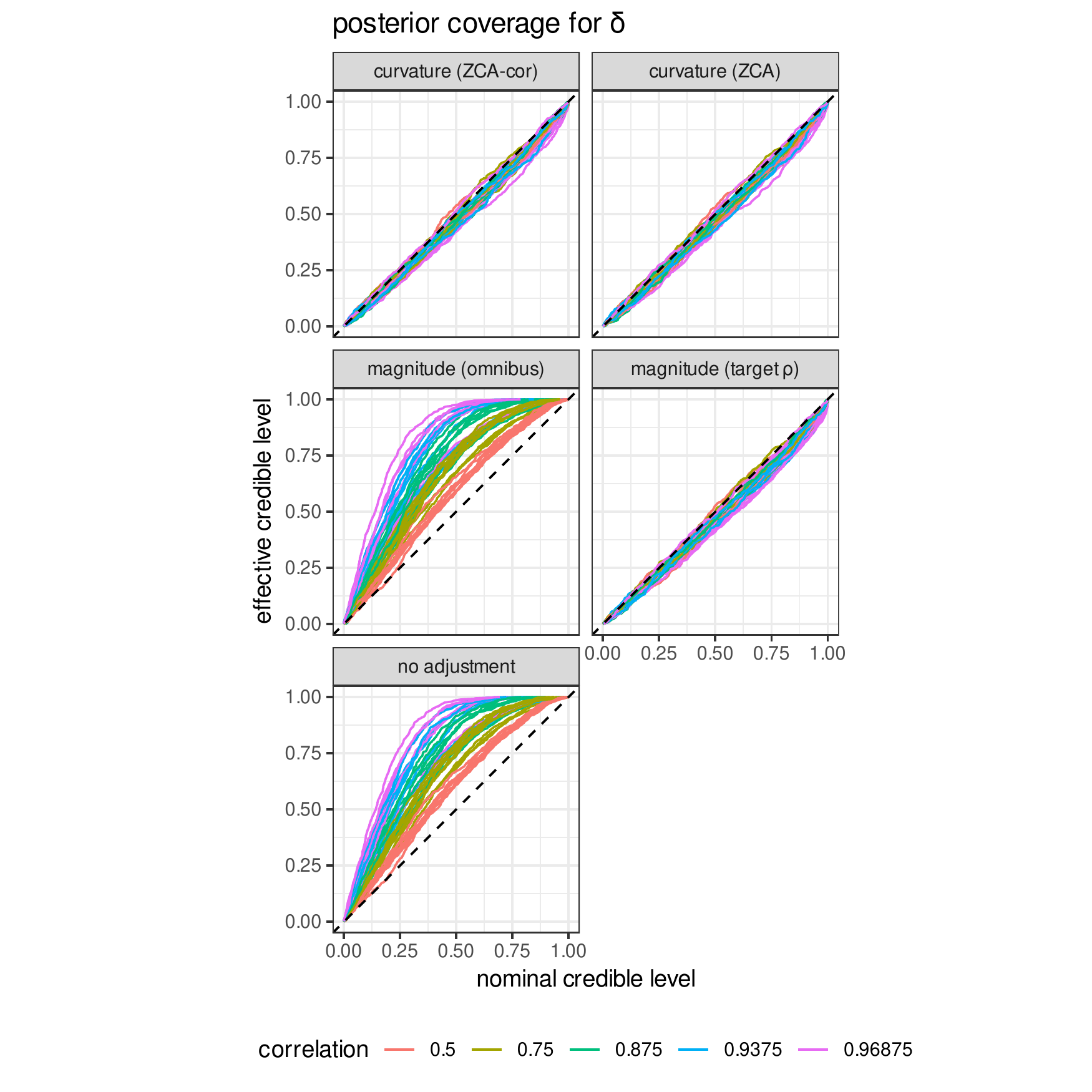}
\end{figure}

At most 1 every 1000 simulations was troubled for either the classical estimation step or the subsequent Bayesian analysis and was discarded. The results reported in the following are thus conditional to no error reported in code. Results are shown in Figures \ref{fig:calib_a1}, \ref{fig:calib_dOR}, \ref{fig:calib_avgdiff}.

The unadjusted posterior here is equivalent to a magnitude adjustment with $w=1$ and it is compared with the standard according to \cite{Ribatet}, which is the curvature method. The standard is presented in two variants, based on the definitions ZCA and ZCA-cor of matrix square roots illustrated by \cite{Kessy}. The magnitude-adjusted posterior is used in both the \textit{omnibus} version by \cite{Pauli} and our variant targeted at $\dOR$. Performance on a second parameter $\avgdiff$ of potential interest is monitored as well.

Each distinct curve in calibration plots summarizes either 1000 analyses from phase 1 or 500 analyses from phase 2, the latter being drawn from $\theta'$ or $\theta''$, while the former cannot be separated according to the ground truth, which is distinct each time by sampling. The curves are colored according to the copula's rank correlation, since this seems to explain most of the variability.

As one can see, there are visible issues in three cases: when the unadjusted posterior or the \textit{omnibus} magnitude adjustment are used to infer $\eta$, and when the magnitude adjustment targeted to $\dOR$ is used to infer a marginal parameter.

As to the first issue about the unadjusted posterior, when making inference on a function $\eta$ of either $(\alpha_1,\beta_1)$ or $(\alpha_2,\beta_2)$ alone, the relevant likelihood is just one of the proper likelihood components, so the unadjusted composite likelihood yields valid inference on $\eta$, but if $\eta$ is a function that involves both groups of parameter components the information is assessed incorrectly.

As to the second issue about the \textit{omnibus} magnitude adjustment, one has to consider that here each marginal distribution is correctly specified and ruled by a different set of parameters, so the matrix $H$ is block-diagonal and its diagonal blocks coincide with the corresponding entries in $J$. So, after a little linear algebra, one can see that the \textit{omnibus} tuning constant is exactly $w=1$ when $J$ is known. The equality holds approximately when $J$ is estimated. This is especially concerning, because the adjustment does not work even under the model.

As to the last issue about the targeted magnitude adjustment, the arguments by \cite{Ribatet} hold, as they stress the unreliability of magnitude adjustments as to anything but the target, see Figure \ref{fig:calib_a1}. Here we stress the fact that magnitude adjustments seem able to recover posterior calibration to essentially the same extent as the curvature method at least as far as the target $\dOR$ is concerned, which is clearly the focus in the problem at hand, see Figure \ref{fig:calib_dOR}.

As a last remark, parameters that are related to the target $\dOR$, such as $\avgdiff$, can benefit from magnitude adjustments even when they are not directly targeted, see Figure \ref{fig:calib_avgdiff}. It is useful to consider the expression for gradients of $\dOR$ and $\avgdiff$: when the average probabilities $\pi_1$ and $\pi_2$ are close to each other, those gradients are proportional and the magnitude adjustment targeted at $\dOR$ and the one targeted at $\avgdiff$ are similar, so one can target either of the two parameters of interest directly, and this will approximately imply targeting the other too.

\section{Discussion}

In this paper we considered some Bayesian adjustment methods for composite likelihoods as reviewed by \cite{Ribatet}. They proposed an adaptive version to these adjustments, along with the version we used for our study: our version is called overall adjustment in their work and it is tuned once before sampling the posterior. Their adaptive version essentially repeats the tuning of constants for each method before each iteration in posterior sampling {\`a}-la Gibbs. We did not use this adaptive version of the adjustments as checking for convergence would have been more difficult. Convergence diagnostics are not reported here, but they were checked in just few cases in order to tune the burn in and thinning behavior of the posterior sampler.

The adjustments discussed here have some intuitive properties that make their tuning rather straightforward. These methods require to perform a frequentist analysis before the Bayesian one. Magnitude adjustments additionally require stating a scalar target parameter, which \textit{a priori} is the only parameter for which the posterior will be approximately calibrated.

We hinted at how the analysis of clustered data with independence likelihoods can benefit from magnitude adjustments in targeting specific parameters of interest. Actually, we managed to make reliable inference even on a parameter that was not directly targeted but still related and potentially relevant. Some known effect measures in medical statistics can approximate each other, so a magnitude adjustment can be useful to a broader extent than the nominal. The additional flexibility of curvature adjustments is not needed in such cases.

The warnings expressed by \cite{Ribatet} still stand, as magnitude adjustments do not recover full posterior calibration, both in the \textit{omnibus} and the targeted version, but more in general under any tuning rule due to intrinsic lack of flexibility. When the scope of the analysis is not limited enough, the more flexible curvature adjustment should be used instead of the \textit{omnibus} magnitude method, which is generally unsafe.

\bibliographystyle{hapalike.bst}
\bibliography{main}

\begin{thebibliography}{}

\bibitem[Chandler and Bate, 2007]{Chandler}
Chandler, R.~E. and Bate, S. (2007).
\newblock Inference for clustered data using the independence loglikelihood.
\newblock {\em Biometrika}, 94:167--183.

\bibitem[Chen et~al., 2015]{Chen}
Chen, Y., Hong, C., Ning, Y., and Su, X. (2015).
\newblock Meta-analysis of studies with bivariate binary outcomes: a marginal
  beta-binomial model approach.
\newblock {\em Statistics in Medicine}, 35:21--40.

\bibitem[Chivers, 2012]{RMHadaptive}
Chivers, C. (2012).
\newblock {\em MHadaptive: {G}eneral {M}arkov {C}hain {M}onte {C}arlo for
  {B}ayesian inference using adaptive {M}etropolis-{H}astings sampling}.
\newblock R package version 1.1-8.

\bibitem[Hofert et~al., 2020]{Rcopula}
Hofert, M., Kojadinovic, I., Maechler, M., and Yan, J. (2020).
\newblock {\em copula: {M}ultivariate dependence with copulas}.
\newblock R package version 1.0-1.

\bibitem[Hong et~al., 2021]{Rxmeta}
Hong, C., Luo, C., Tong, J., and Chen, Y. (2021).
\newblock {\em xmeta: {A} toolbox for multivariate meta-analysis}.
\newblock R package version 1.3-0.

\bibitem[Kessy et~al., 2018]{Kessy}
Kessy, A., Lewin, A., and Strimmer, K. (2018).
\newblock Optimal whitening and decorrelation.
\newblock {\em The American Statistician}, 72:309--314.

\bibitem[Lazar, 2003]{Lazar}
Lazar, N.~A. (2003).
\newblock Bayesian empirical likelihood.
\newblock {\em Biometrika}, 90:319--326.

\bibitem[Miller, 2019]{Miller}
Miller, J.~W. (2019).
\newblock Asymptotic normality, concentration, and coverage of generalized
  posteriors.
\newblock {\em arXiv preprint}, arXiv:1907.09611v1.

\bibitem[Monahan and Boos, 1992]{Monahan}
Monahan, J.~F. and Boos, D.~D. (1992).
\newblock Proper likelihoods for bayesian analysis.
\newblock {\em Biometrika}, 79:271--278.

\bibitem[Müller, 2013]{Muller}
Müller, U.~K. (2013).
\newblock Risk of {B}ayesian inference in misspecified models, and the sandwich
  covariance matrix.
\newblock {\em Econometrica}, 81:1805--1849.

\bibitem[Nelsen, 2007]{Nelsen}
Nelsen, R.~B. (2007).
\newblock {\em An Introduction to Copulas}.
\newblock Springer.

\bibitem[Pauli et~al., 2011]{Pauli}
Pauli, F., Racugno, W., and Ventura, L. (2011).
\newblock Bayesian composite marginal likelihoods.
\newblock {\em Statistica Sinica}, 21:149--164.

\bibitem[Ribatet et~al., 2012]{Ribatet}
Ribatet, M., Cooley, D., and Davison, A.~C. (2012).
\newblock Bayesian inference from composite likelihoods, with an application to
  spatial extremes.
\newblock {\em Statistica Sinica}, 22:813--845.

\bibitem[Shaby, 2014]{Shaby}
Shaby, B.~A. (2014).
\newblock The open-faced sandwich adjustment for {MCMC} using estimating
  functions.
\newblock {\em Journal of Computational and Graphical Statistics}, 23:853--876.

\bibitem[Stoehr and Friel, 2015]{Stoehr}
Stoehr, J. and Friel, N. (2015).
\newblock Calibration of conditional composite likelihood for bayesian
  inference on gibbs random fields.
\newblock {\em arXiv preprint}, arXiv:1502.01997v2.

\bibitem[Tsukahara, 2005]{Tsukahara}
Tsukahara, H. (2005).
\newblock Semiparametric estimation in copula models.
\newblock {\em Canadian Journal of Statistics}, 33:357--375.

\bibitem[Varin et~al., 2011]{Varin}
Varin, C., Reid, N., and Firth, D. (2011).
\newblock An overview of composite likelihood methods.
\newblock {\em Statistica Sinica}, 21:5--42.

\end{thebibliography}

\newpage

\appendix

\section{Plots}

\subsection{Dataset}

A presentation on the dataset taken from \cite{Chen}. Each axis represents one out of two treatment groups, the logarithm of event odd in each group is reported for each study in their meta-analysis. The summary for each group has been estimated based on their marginal beta-binomial approach and the standard error is based on the sandwich covariance matrix.

\newpage

\subsection{Calibration plots}

In the following pages, calibration plots are listed. For each combination of phase, true $\theta$ if phase is 2, true copula and true correlation within study, the empirical cumulative distribution function of statistic $h$ is reported as a distinct line. Each line pools 1000 simulations if phase is 1 and 500 simulations if phase is 2. This is repeated for each adjustment method considered in the study. The plot can be read as effective credibility level (vertically) versus nominal credibility level (horizontally).

Lines are colored according to true correlation, which explains most of the variability when this is present. It is evident that the higher the correlation within studies, the worse the coverage properties of the unadjusted posterior when looking at parameters $\dOR$ and $\avgdiff$, which are functions of all components of $\theta$. In the trivial case where probabilities are independent within studies, the composite likelihood is proper and one may expect all methods to fall back to it.

\newpage

\newpage

\newpage

\end{document}